# Pressure-induced magnetic collapse and metallization of TlFe$_{1.6}$Se$_2$

P. G. Naumov[1,2], K. Filsinger[1], S. I. Shylin[3,4], O. I. Barkalov[1,5], V. Ksenofontov[3], Y. Qi[1], T. Palasyuk[6], W. Schnelle[1], S. A. Medvedev[1], M. Greenblatt[7] and C. Felser[1]

[1] *Max Planck Institute for Chemical Physics of Solids, Nöthnitzer Str. 40, 01187 Dresden, Germany*

[2] *Shubnikov Institute for Crystallography of Federal Scientific Research Centre "Crystallography and Photonics" of Russian Academy of Sciences, Leninskii prospekt 59, Moscow, 119333, Russia*

[3] *Institute for Inorganic and Analytical Chemistry, Johannes Gutenberg University Mainz, 55128 Mainz, Germany*

[4] *Department of Chemistry, Taras Shevchenko National University of Kyiv, 01601 Kyiv, Ukraine*

[5] *Institute for Solid State Physics, Russian Academy of Sciences, Academician Ossipyan str. 2, Chernogolovka, Moscow District, 142432, Russia*

[6] *Institute for Physical Chemistry, Polish Academy of Sciences, PL-01224 Warsaw, Poland*

[7] *Department of Chemistry and Chemical Biology, Rutgers University, Piscataway, NJ 08854, USA*

**Abstract.** The crystal structure, magnetic ordering, and electrical resistivity of TlFe$_{1.6}$Se$_2$ were studied at high pressures. Below ~7 GPa, TlFe$_{1.6}$Se$_2$ is an antiferromagnetically ordered semiconductor with a ThCr$_2$Si$_2$-type structure. The insulator-to-metal transformation observed at a pressure of ~ 7 GPa is accompanied by a loss of magnetic ordering and an isostructural phase transition. In the pressure range ~ 7.5 - 11 GPa a remarkable downturn in resistivity, which resembles a superconducting transition, is observed below 15 K. We discuss this feature as the possible onset of superconductivity originating from a phase separation in a small fraction of the sample in the vicinity of the magnetic transition.

PACS number(s): 74.70.Xa, 74.62.Fj, 75.30.−m

## I. Introduction

The family of iron-based superconductors $A_{1-y}$Fe$_{2-x}$Se$_2$ ($A$=K, Rb, Cs), which adopts a defect variant of the ThCr$_2$Si$_2$ structure, has attracted considerable research interest because of the microscopic coexistence of magnetic ordering and superconductivity at a critical temperature above 30 K [1-4]. It is well established that this coexistence is attributable to a nanoscale phase separation, with magnetism and superconductivity occurring in different spatial regions with different compositions [5-8]. For instance, for Rb$_{0.8}$Fe$_{1.6}$Se$_2$, the superconducting phase is determined as vacancy-free Rb$_{0.3}$Fe$_2$Se$_2$ and thus represents FeSe-layers with a doping of 0.15 electrons per Fe-atom [9]. The antiferromagnetic phase with a stoichiometry close to the nominal composition is a $\sqrt{5}a \times \sqrt{5}a$ superstructure of the $ThCr_2Si_2$ structural type derived from Fe vacancy ordering. Single-phase superconducting samples of this structural type have not been synthesized to date, and thus the intrinsic properties of the superconducting phase remain unclear.

The application of pressure has a strong effect on both the superconductivity and antiferromagnetic ordering [10-12]. Mössbauer spectroscopic studies at pressures above 5 GPa have revealed the appearance of a new paramagnetic phase from the antiferromagnetic phase in $Rb_{0.8}Fe_{1.6}Se_2$ [10]. This new phase might be associated with the reentrant superconductivity at $T_c$=48 K observed at higher pressure (above 10 GPa) [12] in superconducting $A_xFe_{2-y}Se_2$ systems. Contrary to the alkali-metal-containing compounds $A_xFe_{2-y}Se_2$ with their compositional flexibility, isostructural $TlFe_{1.6}Se_2$ can be synthesized only with fully occupied Tl-sites and thus with a limited degree of inherent disorder. As a result, the phase separation known to occur in $A_xFe_{2-y}Se_2$ systems is not observed in insulating antiferromagnetic $TlFe_{1.6}Se_2$ with its ordered Fe-vacancies. Moreover, the compound displays magnetic behavior that differs from that of the alkali-metal analogs [13-15]. Regarding the interplay between the structure, magnetism, and electrical transport properties of Fe-based superconductors, it is of interest to study the pressure effect on the properties of $TlFe_{1.6}Se_2$. In this work, we present the observation of the pressure-induced metallization of $TlFe_{1.6}Se_2$ at a pressure of $\approx$ 7 GPa. The observed metallization is associated with an isostructural phase transition and the collapse of magnetic ordering.

**II. Experimental details**

Single-crystalline samples of $^{57}$Fe-enriched (20 %) $TlFe_{1.6}Se_2$ were prepared by the modified Bridgman method. Thallium rods, $^{57}$Fe-enriched iron powder, and selenium shots were mixed in the ratio 1:1.6:2, placed in an alumina crucible, and sealed in a quartz glass ampoule under an argon atmosphere of 200 mbar. The ampoule was heated to 1343 K and subsequently cooled to 460 K at a rate of 2 K/h. Shiny black plate-like single crystals were obtained with dimensions up to 6 mm. The crystals were structurally characterized by single-crystal X-ray diffraction. Inductively coupled plasma optical emission spectrometry (ICP-OES), energy dispersive X-ray spectroscopy (EDXS), and density measurements were performed to confirm the stoichiometry and homogeneity of the single crystals throughout the bulk. The resistivity in magnetic fields was measured by a low-frequency ac method in a Physical Property Measuring System (PPMS-9), and a Magnetic Property Measuring System (MPMS-XL7) was used for magnetization measurements.

The electrical resistivity at different pressures was measured by the van der Pauw technique in an in-house-designed diamond anvil cell equipped with diamond anvils with a 500 μm culet. The sample was insulated against the tungsten-metal gasket by using a mixture of cubic BN with epoxy.

The sample chamber had a diameter of 200 μm and an initial thickness of 40 μm. The sample was loaded without a pressure-transmitting medium.

The angle-dispersive powder X-ray diffraction studies were performed at room temperature at beamlines 01C2 of the National Synchrotron Radiation Research Centre (NSRRC, Taiwan, wavelength 0.564 Å) and ID09 of the European Synchrotron Radiation Facility (ESRF, France, wavelength 0.4145 Å). The powdered sample of $TlFe_{1.6}Se_2$ was loaded with silicone oil as the pressure-transmitting medium.

In the $^{57}Fe$-Mössbauer pressure studies, spectra were recorded using a $^{57}Co(Rh)$ point source. Small flakes of the single-crystalline sample were placed in a diamond anvil cell (DAC) with silicone oil as the pressure-transmitting medium. Another mosaic Mössbauer absorber measured at ambient pressure was prepared by assembling thin single-crystalline flakes that were separated from the bulk single crystals by the scotch-tape technique. Low-temperature spectra were acquired in transmission geometry using a conventional helium bath cryostat. Mössbauer spectra were recorded in an applied magnetic field at high pressure using the Synchrotron Mössbauer Source at ESRF (ID18). The experimental spectra were fitted using *Recoil* software [16]. Isomer shifts are given relative to *α*-Fe at 295 K. All the samples were prepared in an argon-filled glove box with the $O_2$ and $H_2O$ content below 0.5 ppm.

**III. Results and Discussion**

The stoichiometry of the sample as determined by EDXS and ICP-OES and confirmed by structural refinement of single-crystal X-ray diffraction data was $TlFe_{1.603(1)}Se_{2.018(1)}$ uniformly throughout the crystals. In spite of slow cooling, the vacancies in the studied samples are not completely ordered as indicated by the single-crystal structural refinement (Table I). Accordingly, the temperature dependence of the magnetic susceptibility (Fig. 1a) is similar to that reported for $TlFe_{1.6}Se_2$ with partially ordered Fe-vacancies [15] and shows two magnetic transitions near 100 K and 140 K associated with the canting of the Fe magnetic moments toward the *ab* plane occurring between these temperatures [13, 14]. The electrical resistivity measurements (Fig. 1b) reveal the semiconducting character of $TlFe_{1.6}Se_2$ over the whole temperature range, which is in agreement with previous studies [15]. However, there are no observable anomalies in the resistivity at the temperatures at which the magnetic transitions occur, that is, at 100 K and 140 K, in contrast to those reported previously [15]. Apparently, these weak features are masked by scattering processes that are enhanced owing to the higher level of disorder among the vacancies in our sample.

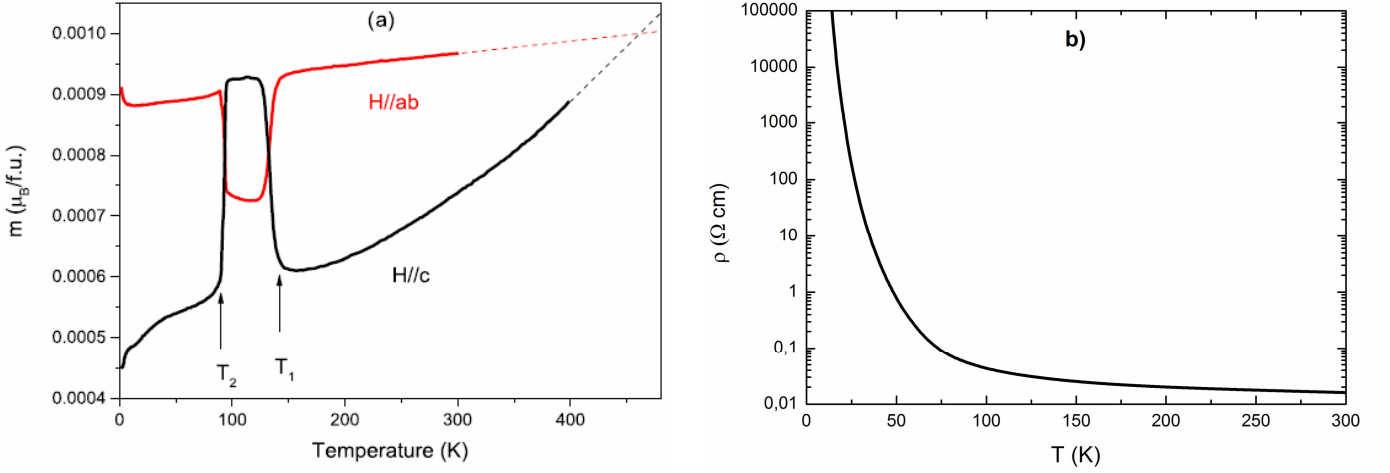

**Fig. 1.** (a) Magnetization of TlFe$_{1.6}$Se$_2$ measured with a magnetic field of 0.1 T applied along the *ab* plane (red) and along the c-axis (black). Extrapolations of the fitted curves intersect at 460 K and thus show the approximate Neel temperature. $T_1$ and $T_2$ mark the temperatures at which spin canting occurs. (b) Electrical resistivity of TlFe$_{1.6}$Se$_2$ as a function of temperature.

**Table 1.** Structural data for TlFe$_{1.6}$Se$_2$ determined from single-crystal X-ray diffraction data. Space group *I*4/*m*, $a = 8.67(2)$ Å, $c = 13.97(2)$ Å.

| Atom | Wyckoff pos. | x | y | z | Occupancy |
| --- | --- | --- | --- | --- | --- |
| Tl 1 | 2a | 0 | 0 | 0 | 1 |
| Tl 2 | 8h | 0.4 | 0.2 | 0 | 1 |
| Fe 1 | 4d | 0.5 | 0 | 0.25 | 0.31(1) |
| Fe 2 | 16i | 0.3 | 0.6 | 0.25 | 0.93(5) |
| Se 1 | 4e | 0 | 0 | 0.356 | 1 |
| Se 2 | 16i | 0.2 | 0.4 | 0.356 | 1 |

The Mössbauer spectrum of a mosaic-crystalline TlFe$_{1.6}$Se$_2$ sample acquired at ambient pressure and room temperature (Fig. 2a) is a single-site sextet with an isomer shift of $\delta = 0.54(1)$ mm/s magnetically split due to antiferromagnetic ordering. The hyperfine structure of the spectrum reflects that the direction of the magnetization vector does not coincide with the main axis of the electric-field gradient (EFG) tensor $V_{zz}$. The angle $\theta_{Bq} = 48(3)°$ between $V_{zz}$ and the magnetic hyperfine field $B_{hf} = 210.7(3)$ kOe derived from the full static Hamiltonian model is close to the angle between $V_{zz}$ and the main crystallographic axis *c* (coinciding with the direction of γ-rays) $\theta_{\gamma q} = 47(3)°$. Therefore, at room temperature the local magnetic moments at the Fe atoms are perpendicular to the *ab* plane, a finding that is in agreement with that of May *et al.* [14]. The small relative intensities of lines 2 and 5 in the Mössbauer spectrum can be considered as a distinctive fingerprint of the out-of-plane orientation of magnetic moments. The spectrum of TlFe$_{1.6}$Se$_2$, loaded in the diamond anvil cell at 0.5 GPa at room temperature (Fig. 2b), is similar to the spectrum at ambient pressure (Fig. 2a). When the mosaic sample is cooled down to 105 K, below the magnetic transition temperature $T_1$, the sextet splits into two magnetic subspectra with relative intensities of 65(3) % and 35(3) % (Fig. 2c). The dominant sextet with $B_{hf}$ = 258.7(1) kOe can be described with the aforementioned $\theta_{Bq}$ and $\theta_{\gamma q}$ values close to 45° and, apparently, corresponds to the Fe sites that are not involved in the magnetic transition shown in Fig. 1a. On the other hand, for the newly formed sextet with $B_{hf}$ = 272.9(3) kOe that corresponds to the newly emerged Fe sites, the angles $\theta_{Bq}$ = 86(1)° and $\theta_{\gamma q}$ = 20(1)° can be derived. Thereby, in the present single-crystalline sample at ambient pressure, the magnetization vector lies in the $ab$ plane at approximately one third of the Fe atoms, when the sample is cooled below $T_1$. However, below $T_2$ their magnetic moments restore their original out-of-plane orientation, as is reflected by the Mössbauer spectrum at 90 K (Fig. 2d) by a single magnetic sextet with $\theta_{Bq} \approx \theta_{\gamma q} \approx$ 45°. Apparently, the Mössbauer results reflect disordering of the vacancy sites in the studied sample, similar to that reported in [14].

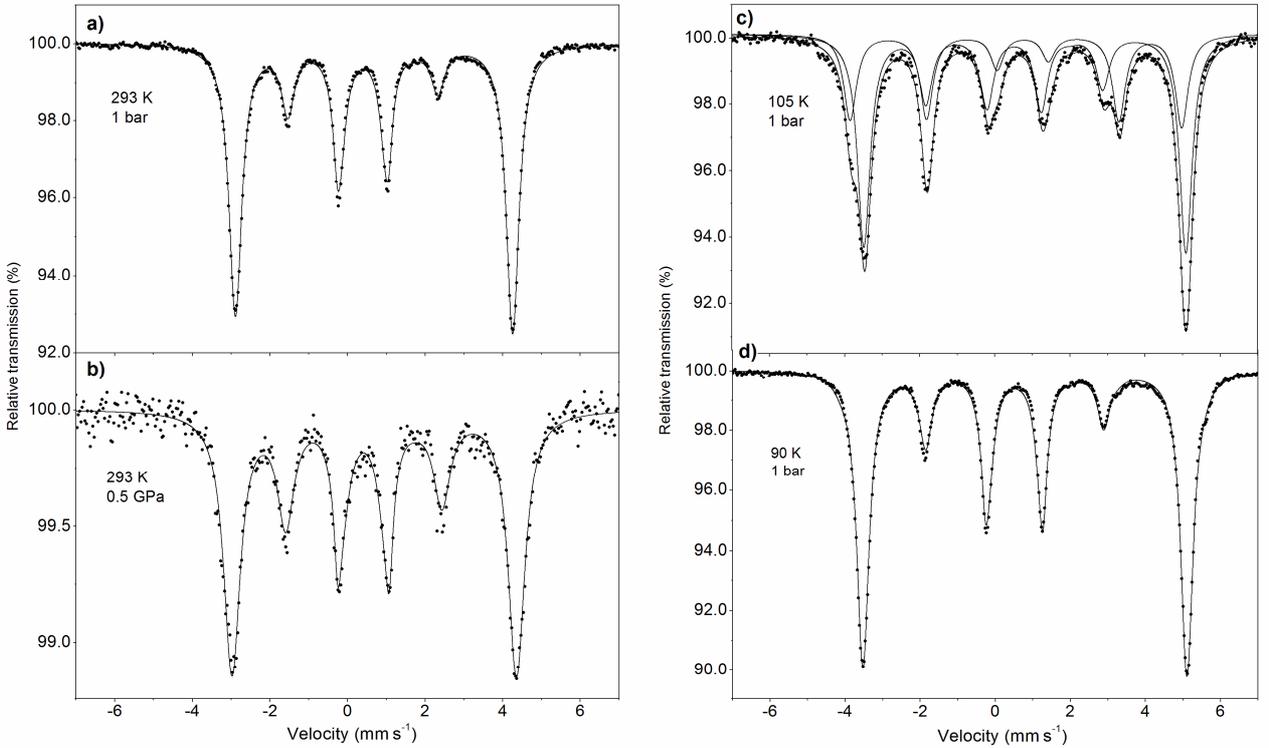

**Fig. 2.** Mössbauer spectra of $^{57}$Fe-enriched TlFe$_{1.6}$Se$_2$ acquired at 293 K from (a) a mosaic-crystalline sample with the $c$-axis parallel to the direction of the γ-rays and (b) the sample in the DAC at an initial pressure of 0.5 GPa. The coexistence of two sextets is observed when mosaic-crystalline TlFe$_{1.6}$Se$_2$ is cooled to 105 K (c) due to partial spin canting between the temperatures $T_1$ and $T_2$ indicated in Fig. 1a. Below $T_2$, only the one Fe-site with parameters close to those above $T_1$ is observed (d).

The evolution of the electrical resistivity (Fig. 3) with increasing pressure shows that, at pressures below 7 GPa, TlFe$_{1.6}$Se$_2$ remains semiconducting, and the X-ray diffraction patterns (Fig. 4) reveal its structural stability in this pressure range. The room-temperature resistivity continuously decreases with increasing pressure (Fig. 3a). The Arrhenius plots (inset in Fig. 3a) demonstrate a

6gradual decrease of the slope, indicating a decrease in the band gap of the semiconducting $TlFe_{1.6}Se_2$ as the pressure increases. In the X-ray diffraction patterns at low pressures (Fig. 4a), all diffraction peaks are shifted towards higher diffraction angles due to the lattice parameters diminishing under pressure (Fig. 4b).

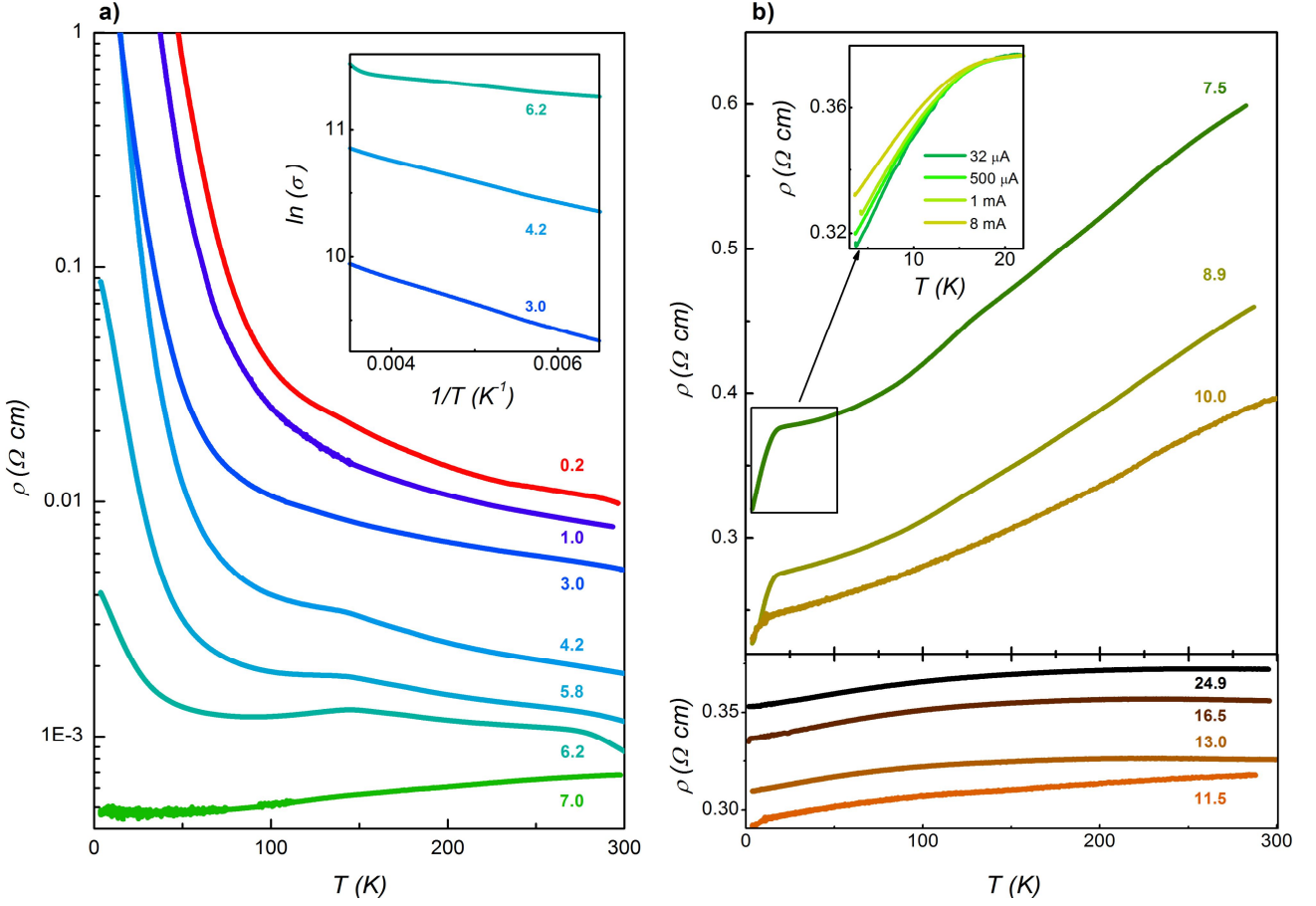

**Fig. 3**. Temperature dependence of the resistivity of $TlFe_{1.6}Se_2$ at pressures below (a) and above (b) 7 GPa. The inset in (a) shows the Arrhenius plots at different pressures. The inset in (b) shows the dependence of low-temperature resistivity measured with different probe currents at 7.5 GPa. The red curve at 0.2 GPa in (a) shows the resistance curve recorded after pressure release from the highest value used in this experiment (25 GPa).

At a pressure of $P_c \approx 7$ GPa the resistivity of the sample continuously decreases with decreasing temperature over the whole temperature range from room temperature down to 4 K, thereby indicating a semiconductor-to-metal transition in $TlFe_{1.6}Se_2$ at this pressure. This is in good agreement with the results of first-principles calculations, which, for $TlFe_{2-x}Se_2$, predict band gap closure to occur at a pressure of ~6 GPa. [17] At a slightly higher pressure of 7.5 GPa, a remarkable kink in the temperature dependence of the resistivity is observed at $T \approx 15$ K followed by a rapid decrease in the resistivity of the sample below this temperature (Fig. 3b). This behavior could be explained by the possible onset of superconductivity in a small fraction of the probed sample below the percolation limit. The shift of the resistivity curve to the low-temperature side below the kink temperature with an increase in the current density (inset in Fig. 3b) supports this idea. However, a state of zero-resistance was not attained when cooling the sample to 4.0 K, indicating that superconductivity may exist only in a small fraction of the sample [18]. This identical behavior was repeatedly observed on five samples from different crystal batches, thereby confirming that this is an intrinsic property of $TlFe_{1.6}Se_2$.

Compared to $A_{1-y}Fe_{2-x}Se_2$ compounds with $A$ = K, Rb, and Cs, $TlFe_{1.6}Se_2$ shows a different structural response during compression. The superconducting $A_{1-y}Fe_{2-x}Se_2$ compounds retain their tetragonal $ThCr_2Si_2$-type structure up to pressures of approximately 19 GPa [10, 19], and only the disappearance of the superstructure ordered on the basis of $\sqrt{5}a \times \sqrt{5}a$ Fe-vacancies is observed. This disappearance is associated with the loss of antiferromagnetic ordering and the suppression of superconductivity without any structural phase transition [12, 19, 20]. In contrast, the X-ray diffraction patterns of $TlFe_{1.6}Se_2$ clearly indicate a structural phase transition as evident by the splitting of the (110) peak at a pressure of 6.8 GPa and the gradual redistribution of the (103) peak intensity to the split (100) peak upon further pressure increase (Fig. 4b). At a pressure of 9.2 GPa the single-phase pattern of the high-pressure phase is obtained (Fig. 4b).

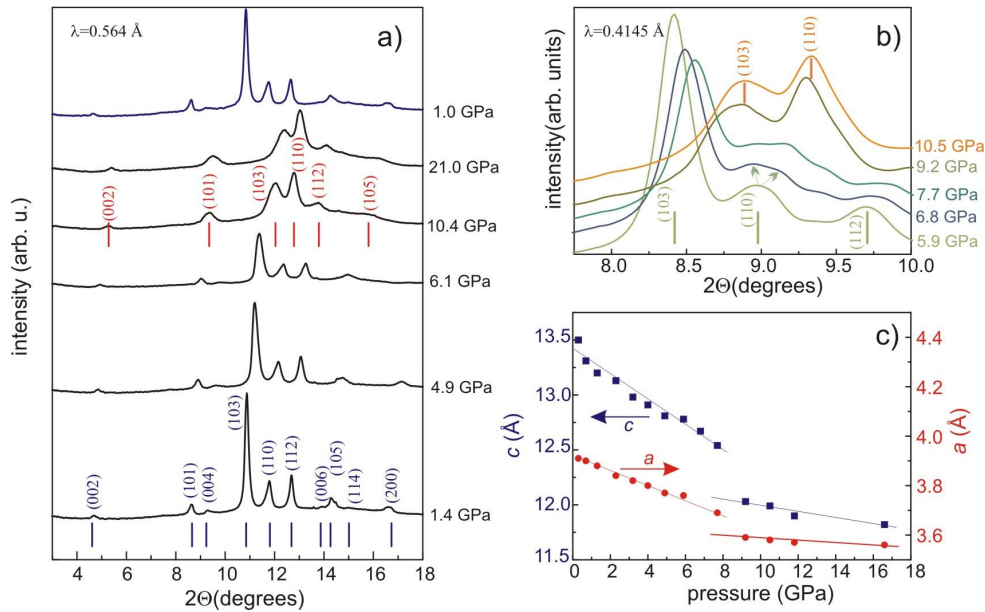

Fig. 4. (*a*) Pressure evolution of the powder X-ray diffraction patterns of $TlFe_{1.6}Se_2$ during pressure increase. The single-phase diffraction pattern at 10.4 GPa is indexed with the $ThCr_2Si_2$-type structure. The blue curve shows the diffraction pattern recorded at 1.0 GPa after pressure release from the highest pressure 21.0 GPa. (*b*) Powder diffraction patterns in the vicinity of the structural phase transition indicated by the splitting of the (110) reflection. (*c*) Decrease of the lattice parameters of $TlFe_{1.6}Se_2$ under pressure showing the collapse of its crystal structure around 8 GPa.

The diffraction pattern of the high-pressure phase can also be properly indexed within the ThCr$_2$Si$_2$-type structure (Fig. 4a), however, with different lattice parameters. This is not surprising since the isostructural phase transition from tetragonal to the so-called collapsed tetragonal structure is a common feature for the 122 family of iron-based superconductors $A$Fe$_2$As$_2$ with the ThCr$_2$Si$_2$-type structure [19-22] and structurally related materials [23, 24]. However, the change in the lattice parameters at the isostructural phase transition in TlFe$_{1.6}$Se$_2$ differs from those observed in most 122 iron-based superconductors. The common structural trend in $A$Fe$_2$As$_2$ compounds is a continuous decrease in the lattice parameter $c$ and an anomalous increase in the parameter $a$ with increasing pressure in the course of the isostructural transformation [19-22]. This resembles the second-order phase transition in compounds with the ThCr$_2$Si$_2$-structure [24]. Contrary to this, both lattice parameters $a$ and $c$ in TlFe$_{1.6}$Se$_2$ decrease discontinuously at the pressure of the phase transition (Fig. 4c), as is typical for a first order phase transition. Possibly, Fe-deficiency results in the $ab$ planes becoming more compressible in TlFe$_{1.6}$Se$_2$ and therefore no anomalous behavior of the $a$-axes is observed.

The transition to the metallic state, decrease in resistivity with $T_{onset}$ = 15 K, and isostructural phase transition are associated with the loss of magnetic ordering in TlFe$_{1.6}$Se$_2$ as follows from the Mössbauer spectra acquired at different pressures (Fig. 5). At pressures below 5.0 GPa the spectra indicate persisting magnetic ordering at both room temperature and 4.2 K. At 7.8 GPa the magnetic sextet coexists with the paramagnetic spectrum. It should be emphasized that no structural transition is observed below 7 GPa as evidenced from powder diffraction measurements. At 7.8 GPa the fraction of the paramagnetic component approximates 50 %, and at 9.5 GPa, as well as at higher pressures up to 25.6 GPa, the sample is completely non-magnetic. A comparison of the Mössbauer spectra at different pressures at room temperature and 4.2 K requires the relative intensities of lines 2 and 5 to be taken into account, because, as mentioned above, these lines are the fingerprints of the orientation of magnetic moments at Fe atoms relative to the crystallographic axes. At 4.2 K and at pressures above 2.2 GPa, the relative intensities of lines 2 and 5 attain their maximal value, whereas at 293 K their intensities increase with pressure. This means that the magnetic moments at Fe atoms at room temperature tilt towards the $ab$ plane as the pressure increases, whereas they are already in plane at 4.2 K at approximately 2.2 GPa. The phase with its magnetic moments in the $ab$ plane exists in the 100 K – 140 K temperature range at ambient pressure, however under pressure it stabilizes over a much broader range. In other words, pressure favors the magnetic ground state with the in-plane magnetic moments at the Fe atoms similar to those that were previously observed in samples with completely ordered vacancies [14].

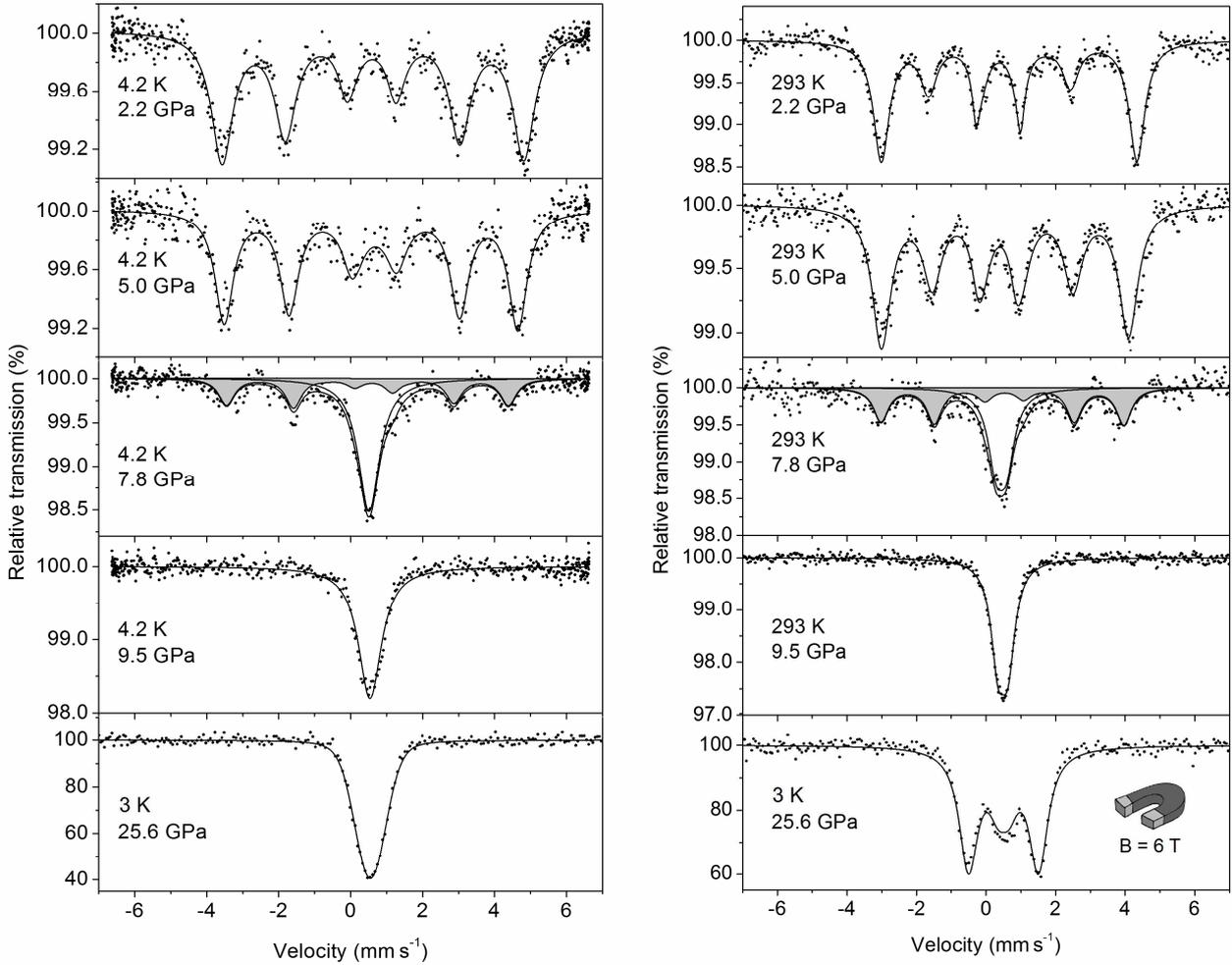

**Fig. 5.** Pressure-dependent Mössbauer spectra of $^{57}$Fe-enriched TlFe$_{1.6}$Se$_2$ at room and low temperature showing antiferromagnetic collapse. Coexistence of the magnetic and paramagnetic states in approximately equal amounts is observed at 7.8 GPa. Application of the magnetic field of 6 T to the sample at 25.6 GPa magnetically splits the Mössbauer spectrum. Fitting of the in-field spectrum with an effective hyperfine magnetic field of 6.0(2) T indicates no localized magnetic moments at the Fe atoms in the metallic phase of TlFe$_{1.6}$Se$_2$.

The appearance of a paramagnetic phase at pressures above 7 GPa with hyperfine parameters typical for Fe selenides [25] indicates collapse of the antiferromagnetic structure with in-plane magnetic moments at the Fe atoms. The observed decrease in resistivity with $T_{onset}$ = 15 K, which is associated with the collapse of the antiferromagnetic ordering in this case, resembles that observed in the related $A_{1-y}$Fe$_{2-x}$Se$_2$ compounds, where superconductivity re-emerged at ~11 GPa [12]. Above 11 GPa, the compound is completely paramagnetic with no evidence of intrinsic static or fluctuating dynamic magnetic moments as confirmed by the Mössbauer spectroscopy experiment in an external magnetic field of 6 T at 25.6 GPa (Fig. 5). Release of pressure completely restores original magnetically-split structure of Mössbauer spectra confirming the mentioned above reversibility of the structure.

The possible onset of superconductivity in TlFe$_{1.6}$Se$_2$ at a pressure slightly above the isostructural transition is unexpected, because for 122 Fe-superconductors, a pressure-induced transition in the collapsed tetragonal phase was proposed to account for the suppression of antiferromagnetic spin fluctuations and consequent disappearance of superconductivity at high pressure [26]. It might be speculated that, in the case of TlFe$_{1.6}$Se$_2$, structural modifications lead to the suppression of antiferromagnetism and the system experiences a narrow spin-fluctuating state. As a result, the onset of superconductivity is observed only in a relatively narrow pressure range above the structural transition, although zero resistance is not reached. However, the most probable explanation would have to consider the dual character of the influence of the disappearance of the magnetic order in TlFe$_{1.6}$Se$_2$ upon attaining superconductivity. On the one hand, the coexistence of AFM clusters with emerging paramagnetic regions could be considered a microphase separation that favors superconductivity similar to the aforementioned A$_x$Fe$_{2-y}$Se$_2$ (A = K, Rb, Cs) systems. On the other hand, the simultaneous appearance under pressure of non-compensated magnetic moments of several Bohr magnetrons tends to destroy the superconducting state similar to the scenario observed in the pressure study of Rb$_{0.8}$Fe$_{1.6}$Se$_2$ [10]. The current observation makes it possible to conclude that the latter mechanism dominates the competition. Therefore, the appearance of bulk superconductivity during the decay of the magnetically ordered state in TlFe$_{1.6}$Se$_2$, even in a narrow pressure range, could hardly be expected. Most probably, the observed superconductivity is of a filamentary nature in a small sample fraction and could be detected by resistivity measurements as opposed to volume-sensitive measurements such as magnetic susceptibility.

**IV. Conclusions**

The structural, magnetic, and electrical transport properties of TlFe$_{1.6}$Se$_2$ were studied at high pressure by means of synchrotron X-ray powder diffraction, Mössbauer spectroscopy, and electrical resistivity measurements. The structural characterization of the as-grown single crystals at ambient pressure reveals the crystal structure to be a $\sqrt{5}a \times \sqrt{5}a$ supercell of the ThCr$_2$Si$_2$-type structure with partial ordering of the Fe-vacancies. At ambient conditions, TlFe$_{1.6}$Se$_2$ is an antiferromagnetically ordered semiconductor. At pressures below 7 GPa, no changes in the structure, magnetic ordering, and electrical transport were observed. However, at ~7 GPa our electrical resistivity measurements indicated an insulator–metal transition. The observed change in the electrical properties is associated with a suppression of the magnetic ordering, as revealed by Mössbauer spectroscopy. At the same pressure, the powder X-ray diffraction patterns indicate a structural phase transition to the collapsed tetragonal phase. In the pressure range 7.5–11 GPa, a clear precipitous decrease in the resistivity below $T \approx 15$ K is observed. This decrease might be associated with the onset of superconductivity in a small fraction of the sample in the vicinity of the antiferromagnetic-to-paramagnetic transition where phase separation is the most probable. Although zero resistance is not reached, this observation supports the idea of an unconventional mechanism of superconductivity in iron chalcogenides, where magnetic fluctuations play a key role in superconducting pairing [27]. The presented results show that $TlFe_{1.6}Se_2$ is a distinctive example of interrelation between the structure, magnetism, and superconductivity in the family of Fe-based superconductors.

**Acknowledgements**

This work was supported by the DFG within Priority Program No. 1458 by Grants ME 3652/1-2 and KS51/2-2. We acknowledge the ESRF and NSRRC for granting the beam time, and we are grateful to Michael Hanfland for providing assistance in using beamline ID09, Alexander Chumakov and Rudolf Rüffer for their help with SMS measurements at ID18, and Hwo-Shuenn Sheu of NSRRC for assistance in using beamline 01C2. We are thankful for the experimental assistance provided by Ralf Koban. MG acknowledges NSF-DMR grant 1507252. TP gratefully acknowledges support from the Polish National Science Centre within project 2012/05/E/ST3/02510 ("SONATA BIS"). PN acknowledges the support by the Russian Scientific Foundation (Project No. 17-72-20200).